\documentclass[aps,pra,twocolumn,showpacs]{revtex4}
\usepackage{graphics}
\usepackage{dcolumn}
\usepackage{epstopdf}
\usepackage{epsfig}

\newcommand{\Xstate}{\mbox{X$^1\Sigma_{\mathrm{g}}^+$}}
\newcommand{\Astate}{\mbox{A$^1\Sigma_{\mathrm{u}}^+$}}
\newcommand{\Bstate}{\mbox{(1)$^1\Pi_{\mathrm{u}}$}}

\begin{document}

\title{The $A^{1}\Sigma^{+}_{u}$ + (1)$^1\Pi_u$ system of Mg$_2$}

\author{Horst Kn\"ockel, Steffen R\"uhmann, Eberhard Tiemann } 

\affiliation{QUEST and Institut f\"ur Quantenoptik, Leibniz Universit\"at Hannover, Welfengarten 1,
30167 Hannover, Germany}

\date{\today}

\begin{abstract}
The \Astate~- \Xstate~UV spectrum of Mg$_2$ has been investigated with high resolution employing Fourier-transform spectroscopy and laser excitation. Computer simulation and fit of line positions to the overlapping structures in the spectra yield precise transition frequencies. Starting with the well characterized ground state $X^{1}\Sigma^{+}_{g}$ from former work we derived excited energy levels and report on the evaluation of the $A^{1}\Sigma^{+}_{u}$ excited state, which is found to interact with another electronic state, which we identify as the lower part of the (1)$^1\Pi_u$ state. A coupled channels fit to the level energies of the upper state yields a reliable potential energy curve for the \Astate~ for the range of vibrational levels 1~$\le$~$v'$~$\le$~46. A potential energy curve for the (1)$^1\Pi_u$ state is proposed, but the (1)$^1\Pi_u$ state is only characterized by its coupling to the A state, and no direct transition to a level of the (1)$^1\Pi_u$ could be uniquely identified due to the overlapping spectral structures.
\end{abstract}


\maketitle


\section{Introduction}
\label{intro}

Recently we reported on new investigations of the Mg$_2$ UV - absorption spectrum \cite{Xpaper}. Major motivation of the experimental efforts was a reliable characterization of the cold collision properties of ground state atoms by the molecular ground state \Xstate, which were derived from the long range behaviour of the potential energy curve (PEC) and given in the form of scattering lengths. With the new data, incorporating also data from former spectroscopic work \cite{Balfour,Stwalley1,Stwalley2,Stwalley3,Scheingr,Vidal} as much as possible, paper \cite{Xpaper} exclusively handles the properties of the molecular ground state and concentrates on different mathematical forms to represent the PEC. The 
spectroscopic data on the upper state level structure is more complex, and its evaluation will be reported in the present paper. Due to the better precision compared to \cite{Balfour} and the computer aided evaluation of the spectral structures, perturbations manifesting themselves by shifts of the observed spectral lines could be identified.  The (1)$^1\Pi_u$ state was included as the most plausible perturber in the modeling of the excited state data to derive the PEC for state \Astate. 

The most abundant isotopes of the alkaline earth metals have nuclear spin zero, thus there is no hyperfine structure, a fact which simplifies the spectral structures. Similar investigations have been carried out in the past on Ca$_2$ and Sr$_2$ for the ground states \cite{CaX,SrX} and for some of the lowest excited electronic states \cite{Caex,Srex}. 
Generally, all those examples revealed substantial differences between PECs from up to date \textit{ab initio} calculations and PECs deduced from spectroscopic observations. 

\begin{figure}
\centering
\resizebox{0.5\textwidth}{!}{%
 \includegraphics{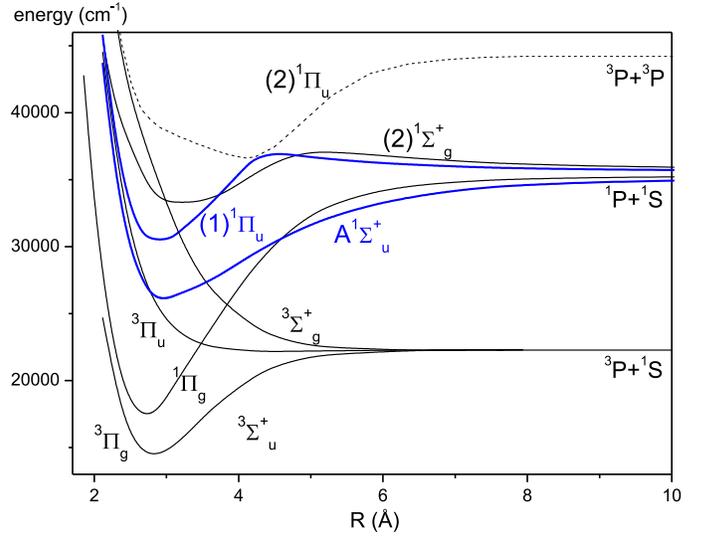} 
}
  \caption{(color online) The lowest excited states of Mg$_2$ according to
  \textit{ab initio} calculations by Amaran et al. \cite{ai-pots}. Thick blue lines indicate
  the PECs of the excited states relevant for this work.} 
  \label{fig:1}
\end{figure}

An overview of the electronic structure of the Mg$_2$ molecule is given in Figure \ref{fig:1}. The data of the PECs have been taken from Amaran et al. \cite{ai-pots}. The PEC of the ground state \Xstate~has been omitted for an enlarged view of the excited states. The thick blue curves are the PECs which are relevant in this report. The \Astate~ state, on which we focus in this paper, is crossed by a $^3\Pi_u$ state from below, which comes from the asymptotic combination of Mg ($^3P$) and Mg($^1S$). A coupling of this state to the A state by spin-orbit coupling will cause predissociation, but this was not identified within our recorded spectra. The inner branch of the \Astate~state potential is close to that of the (1)$^1\Pi_u$ state, in former \textit{ab initio} calculations by Czuchaj et al. \cite{Old-ai} a crossing of both PECs was predicted. We propose below the \Bstate~state as being responsible for the observed perturbations in the A - X spectrum. The crossings of the $^3\Sigma^+_g$ and $^1\Pi_g$~play no role in the present work because of their gerade symmetry.

In the next section we will give a short overview of the relevant 
experimental details. Then we will describe the evaluation of the spectra and the treatment
of the data. In section \ref{modeling} the model approaches will be presented. The paper
ends with a discussion of the results and a short conclusion.

\section{Experimental methods}
\label{expt}

The experimental methods employed are described in detail in \cite{Xpaper}. The Mg$_2$ vapour was generated in a stainless steel heat pipe containing Mg metal at a typical operating temperature of the central part of 1100~K, with about 40 to 80~hPa Ar as buffer gas. Absorption spectra  were taken with a broad band light source, reducing the bandwidth to between 400~nm and 250~nm by optical filtering. A high resolution Fourier-transform spectrometer (FTS) with a typical resolution of about 0.1~cm$^{-1}$, which corresponds well to the Doppler width of the spectral lines, was employed to obtain the spectra. About 300 interferograms were finally averaged to obtain the absorption spectrum actually used for evaluation. In addition, laser induced fluorescence was recorded, excited by an UV Ar-ion laser or a frequency doubled TiSa laser, respectively, employing the FTS for the dispersion of the fluorescence. The  fluorescence progressions having in common one excited level offer substantial information on the lower state and assign clearly the rotational quantum number of the excited level which helps significantly in the assignment of the dense absorption spectrum. Characteristic examples of recorded spectra of these kinds were given in \cite{Xpaper}.

\section{Evaluation of spectra and data treatment}
\label{data}

For the assignment of the spectral lines we found complete agreement to the previous work by Balfour and Douglas \cite{Balfour}. However, due to the isotopic abundance of natural Mg ($^{24}$Mg about 79~\%, $^{25}$Mg 
about 10~\%, $^{26}$Mg about  11~\%), the spectra are a mixture of six isotopologues in the 
molecular gas with abundances of 62~\%, 7.9~\%, 8.7~\%, 1.1~\%, 1.0~\% and 1.2~\% for $^{24}$Mg$_2$, 
$^{24}$Mg$^{25}$Mg, $^{24}$Mg$^{26}$Mg, $^{25}$Mg$^{26}$Mg, $^{25}$Mg$_2$, $^{26}$Mg$_2$, respectively.
The three most abundant isotopic forms could be finally identified in our spectra. For the temperatures of 1100 K necessary in the experiment, all rovibrational levels of the ground state have a significant thermal population due to the small vibrational frequency and the large equilibrium internuclear distance of a typical van der Waals state. Thus the absorption spectrum shows high density of lines and essentially no observed spectral feature is due to a single line, but a blend of lines.

This limitation in unraveling the spectra could partly be overcome by using a computer program, which simulates the spectra and can also fit an artificial spectrum to a selected window of the observed spectrum. It is based on calculating eigenvalues from intermediate PECs, representing the transition frequencies as differences between eigenvalues of upper and lower electronic states and applying standard line profiles for simulation of the spectrum in the selected window. In this way the overlap of lines could be accounted for, at least for the strong lines and for the three most abundant isotopologues $^{24}$Mg$_2$, $^{24}$Mg$^{25}$Mg and $^{24}$Mg$^{26}$Mg. The abundance of the other isotopologues is small enough to neglect their influence within the signal-to-noise ratio of the spectra. Due to the very flat and wide ground state potential, the intensities of the spectral lines not only depend on thermal population of the ground state levels, but also on the significant change of the Franck-Condon factors with rotation. As a consequence nearly no vibrational band ($v'$~-~$v''$) can be observed over the full span of rotational quantum numbers, but always fragments with a certain limited range of rotational quantum numbers appear.

An iterative procedure was pursued increasing the range of quantum numbers of assigned spectral lines and improving the intermediate PECs step by step. We could replace most of the lines assigned in \cite{Balfour} by new frequencies being more precise by about a factor of three. By the evaluation procedure it was also possible 
to identify parts in bands where the lines were shifted with respect to their expected positions, unveiling regions of perturbations in the upper state by a yet unknown electronic state. We note that in paper \cite{Xpaper} we limited the data set to lines, which were obviously not perturbed, restricting  the range of vibrational quantum numbers v~$\le$~28 in the upper state \Astate~for getting a precise description of the ground state. We used the PECs determined with this set of data for lower and upper state to assign lines with higher vibrational quantum numbers $v'$. 
 
When the difference between frequencies of newly assigned lines and their predictions increased by more than a typical full width at half maximum (FWHM, about 0.22 cm$^{-1}$), a new PEC was determined now including those new lines. With this iterative procedure we could benefit from better prediction qualities due to the stiffness 
of the potential representation compared to a Dunham parameter approach and could extend the span of vibrational levels in the upper state beyond $v'$ = 46. During the assignment process, we also identified various regions, where the lines were shifted systematically with respect to the prediction. The lines starting from lower J are first always higher in energy up to a maximum shift at certain J, then switch sign to negative shift and reduce in magnitude when going to higher J. This is a clear signature of a perturbation by coupling to another level system which has a larger rotational constant than the A state. So we identified these regions as being perturbed, as soon as the shifts were larger in magnitude than about 0.06 cm$^{-1}$. The shifts could be verified by finding other lines with similar shifts connecting to the same upper level. Besides the main isotopologue $^{24}$Mg$_2$, only few lines of mixed isotopes could be sufficiently precisely determined due to their mostly low intensities and significant spectral overlap. Therefore, the main body of data for other isotopes stems from the absorption measurements on a sample of enriched $^{26}$Mg$_2$ by Balfour and Douglas \cite{Balfour}.

Above $v'$ = 46 levels lines with additional perturbations were found which indicates the existence of an additional perturber and hinders unique assignment. Therefore, the data set used for the modeling presented here comprises only lines with 1~$\le$~$v'$~$\le$~46. Transitions to $v'$ = 0 were not found, this failure could be finally confirmed by the prediction of small Franck-Condon factors.

Generally, for each excited level we determined at least the P line and the R line of the most prominent vibrational band in order to be sure about having identified the proper line in the dense spectrum. In some cases also lines of less intense bands were used in addition for verification, as far as such lines were intense enough for unique assignment. In the perturbed regions no extra line could be identified with certainty, so for the perturbing state there are no direct observations. In total, we used 5844 lines and their uncertainties range from 0.015 to 0.3~cm$^{-1}$ depending on the signal-to-noise ratio and the degree of overlap of lines. The data from \cite{Balfour} have an uncertainty of 0.05~cm$^{-1}$. The whole data set is available in the electronic supplement \cite{suppl}.

\begin{figure}
\resizebox{0.55\textwidth}{!}{%
  \includegraphics{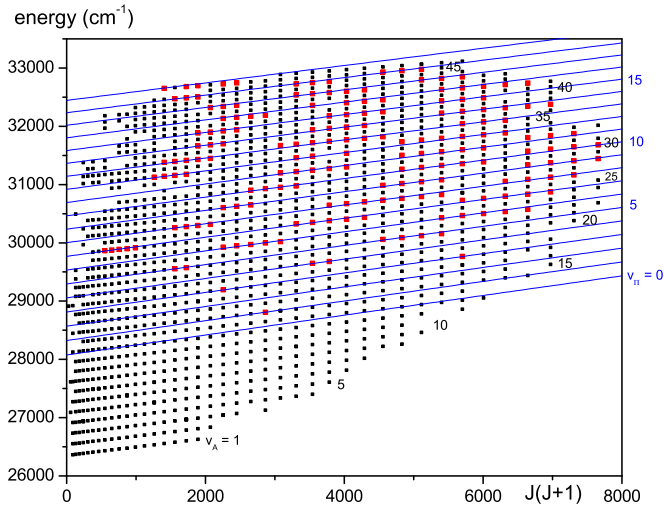} }
  \caption{(color online) Overview of the data set used for the modeling. The squares mark the levels, the black ones are perturbation free, the red ones identify regions where deviations of the levels from a  single state model appear. The full blue lines indicate the course of energy levels of the perturbing \Bstate~state. On the right vibrational quantum numbers are given for orientation.} 
  \label{fig:2}
\end{figure}

\section{Modeling of the excited state A$^1\Sigma^+_u$}
\label{modeling}

Because the ground state is well known from \cite{Xpaper}, we calculate term energies of the upper state \Astate~by adding the energy of the ground state level, which results as eigenvalue of the analytical potential given in table I of \cite{Xpaper}, to the measured transition frequency. All energies are referred to the minimum of the electronic ground state \Xstate. An overview of the data available is given in figure \ref{fig:2}. The full squares mark the level energies in the vibrational interval 1~$\le$~$v'$~$\le$~46. The larger (red) squares mark regions, where levels were identified as being perturbed using only a single state fit of the \Astate~state. The full lines mark the course of energy levels of the later proposed perturbing electronic state assigned as (1)$^1\Pi_u$ (see section \ref{coupled}).                   

\subsection{Construction of the potential energy curves}
\label{constr}

In a first approach the experimentally determined level energies are used for a direct fit of a PEC for 
\Astate~neglecting the perturbations. The radial Schr\"odinger equation with an effective Hamiltonian (e.g. \cite{iod08}) for an electronic state with symmetry $^1\Sigma$ of the form 
\begin{eqnarray}
\bf{H}_{eff}& = & 
 -\frac{\hbar^2}{2\mu}\frac{\partial^2}{\partial R^2}+U(R)+U_{corr}(R) \nonumber\\
 & &+\frac{\hbar^2\cdot [1+\alpha(R)]\cdot J(J+1)}{2\mu R^2}
\label{eq:5}  
\end{eqnarray}
is solved to find eigenvalues; $\mu$ is the reduced mass and $\hbar$ is Planck's constant. U(R) is the Born-Oppenheimer potential and U$_{corr}$ and $\alpha$ adiabatic and non-adiabatic Born-Oppenheimer corrections (BOC), respectively, which could become determinable if more than one isotopologue or high rotational quantum numbers 
are involved. Details are given in \cite{iod08}. The adiabatic correction U$_{corr}$(R) is mass dependent and can only be distinguished from U(R) if data of different isotopic species of the molecule are available. It had to be taken into account for the \Astate~state, as was already mentioned by Vidal et al. \cite{Vidal}. The non-adiabatic BOC was not significant in the present evaluation despite rotational quantum numbers J up to 87. During the fit the eigenvalues for the levels are calculated with the current potential. The difference of the calculated energy to the experimentally determined one is used in a weighted least squares fit in which the parameters of the PEC representation are adjusted to yield a minimum $\chi^2$
\begin{equation}
\chi^2=\sum_{j} \left( \frac{\mathrm{obs}_j-\mathrm{cal}_j}{\Delta \mathrm{obs }_j} \right)^2,
\label{eq:81}
\end{equation}
where  $\mathrm{obs}_j$ are the observed energies,  $\mathrm{cal}_j$ the calculated values and $\Delta  \mathrm{obs}_j$ the estimated uncertainties of the experimental values. For the fitting of the nonlinear problem we use the MINUIT program package \cite{Min90}. 

For the representation of the PEC of the \Astate~state the "X-representation" is used. The PEC U(R) (see e.g. \cite{Srex}) is cut according to physical considerations at the transition points R$_i$ and R$_o$ into three parts. 
The central well around the potential minimum, labeled here by CP, with $R_i \le R \le R_o$  is represented as a finite power series 
\begin{equation}
\label{uanal}
\mbox{U}^{\mathrm {CP}}(R)=T_m+\sum_{i=1}^{n}a_i\,x(R)^i
\end{equation}
with a nonlinear variable function $x$ of internuclear separation $R$:
\begin{equation}
\label{xv}
x(R)=\frac{R - R_m}{R + b\,R_m}.
\end{equation}
\noindent The parameters \{a$_i$\} and T$_m$ are adjusted in the fit. The parameter $b$ determines the pole
at small $R < R_i$, and allows to some degree to account for the asymmetry of the potential, and 
$R_m$ is the expansion center, chosen typically close to the equilibrium internuclear distance 
R$_e$. Those two parameters are found once by initially modeling a preliminary RKR potential with 
the analytic form (\ref{uanal}) and kept fixed in subsequent fits of the parameters a$_i$. 

The potential is continuously and differentiably extrapolated  for  $R < R_{i}$ with:

\begin{equation}
\label{rep}
  \mbox{U}^{\mathrm {SR}}(R)= A + B/R^s     
\end{equation}
\noindent by selecting $s$ and adjusting the parameters $A$, $B$ accordingly at R$_i$.

For large internuclear distances $R > R_{o}$ the standard long range form of molecular potentials 
is adopted:
\begin{equation}
\label{lrexp}
  U^{\mathrm {LR}}(R)=diss.asympt.-\sum_{i} C_i/R^i 
\end{equation}

For the \Astate~of Mg$_2$ terms with $i=3,6,8$ were used. C$_3$ was calculated from the atomic lifetime 
of 2.09(10)~ns reported in \cite{Lundin} for $^1$P$_1$ of Mg, C$_6$ and C$_8$ were determined by the 
requirement of a fixed dissociation asymptote and a continuous and differentiable transition at R$_o$. The dissociation asymptote was set to 35481.745 cm$^{-1}$ with respect to the minimum of the ground state 
using D$_e$(X) = 430.472(500) cm$^{-1}$ \cite{Xpaper} and the level energy E = 35051.27211(17) cm$^{-1}$ of 
Mg $^1$P$_1$ \cite{isv}.

The adiabatic Born-Oppenheimer correction potential U$_{corr}$(R) is represented as
\begin{equation}
U_{corr}(R) = (1-\frac{\mu_{ref}}{\mu}) \cdot U_{ad}
\label{eq:7}
\end{equation}
with
\begin{equation}
U_{ad} = (\frac{2R_m}{R+R_m})^l \sum_{i} v_i \cdot x(R)^i , i=0,1,2,...   
\label{eq:8}
\end{equation} 
with $\mu_{ref}$ being the reduced mass of the selected reference isotope $^{24}$Mg$_2$ and $l$ being the power of R in the leading term of the long range interactions. The atomic isotope shifts of the $^1$P$_1$ level ($\nu(^{25}Mg)-\nu(^{24}Mg)$ = 0.02482(25)~cm$^{-1}$, $\nu(^{26}Mg)-\nu(^{24}Mg)$ = 0.04716(25)~cm$^{-1}$ \cite{isv}) were neglected, because the energy gap between the last observed levels and the Mg($^1$S$_0$) + Mg($^1$P$_1$) asymptote is more than 2500 cm$^{-1}$, large enough that finer details at the asymptote do not influence the results.

\subsection{Results of the single state approach}
\label{single}

Excluding from the full set of 5844 levels all those, which were identified as perturbed, altogether 
N = 5390 transitions remained, 4955 of them for the main isotopologue $^{24}$Mg$_2$ representing 
1625 levels of the \Astate~state. We adjusted T$_m$, 15 potential parameters a$_i$ and one parameter 
v$_0$ for the adiabatic Born-Oppenheimer correction. The final standard deviation was 
0.025 cm$^{-1}$, which corresponds well to the average experimental uncertainty, yielding a reduced 
standard deviation $\sigma=\sqrt{\chi^2/(N-N_p)}$ of $\sigma$ = 0.97, where $N_p$ is the number of 
fitted parameters, 17 in the present case. The results of the fit are listed in table \ref{tab1}.

\begin{table}
	\caption{Parameters for the potential curve of the \Astate~excited state of Mg$_2$. For a faithful representation of the potential the coefficients are given to 10 figures. This does not reflect the accuracy of potential parameters in reproducing experimental energies within the standard deviation.}
	\label{tab1}
		\begin{tabular}{ll}
		\hline \noalign{\smallskip}
			Parameter & Value\\
		\hline \noalign{\smallskip}
		\hline \noalign{\smallskip}
			\multicolumn{2}{l}{$R<R_{i}=$2.40\AA}	 \\ 
			A (cm$^{-1}$)					& 0.2337728514$\cdot10^5$		$^b$	\\
			B (cm$^{-1}$\AA$^s$) 	& 0.1670102671$\cdot10^7$		$^b$ \\
			s             				& 6.0 \\
		\hline \noalign{\smallskip}
			\multicolumn{2}{l}{$R_{i}\leq R \leq R_{o}=$7.00\AA}		\\ 
			$R_{m}$ (\AA)					& 3.0810			\\
			$T_m$   (cm$^{-1}$)   & 26068.9374 \\
			b							    & -0.50				\\
			$a_{1}$  (cm$^{-1}$)			& -0.3184844904$\cdot10^{2}$\\
			$a_{2}$  (cm$^{-1}$)			&  0.1539171479$\cdot10^{5}$\\
			$a_{3}$  (cm$^{-1}$)			&  0.9143888458$\cdot10^{4}$\\
			$a_{4}$  (cm$^{-1}$)			& -0.1208749050$\cdot10^{4}$\\
			$a_{5}$  (cm$^{-1}$)			& -0.1658325000$\cdot10^{5}$\\
			$a_{6}$  (cm$^{-1}$)			&  0.4992091056$\cdot10^{4}$\\
			$a_{7}$  (cm$^{-1}$)			&  0.8728631208$\cdot10^{5}$\\
			$a_{8}$  (cm$^{-1}$)			& -0.6922598796$\cdot10^{5}$\\
			$a_{9}$  (cm$^{-1}$)			& -0.4052591883$\cdot10^{6}$\\
			$a_{10}$ (cm$^{-1}$)			&  0.2193592836$\cdot10^{6}$\\
			$a_{11}$ (cm$^{-1}$)			&  0.9904222700$\cdot10^{6}$\\
			$a_{12}$ (cm$^{-1}$)			& -0.4097327102$\cdot10^{6}$\\
			$a_{13}$ (cm$^{-1}$)			& -0.1327033049$\cdot10^{7}$\\
			$a_{14}$ (cm$^{-1}$)			&  0.3102998712$\cdot10^{6}$\\	
			$a_{15}$ (cm$^{-1}$)			&  0.7591975582$\cdot10^{6}$\\	
		\hline \noalign{\smallskip}
			\multicolumn{2}{l}{$R_{o}<R$}	          \\ 
			$C_{3}$ (cm$^{-1}$\AA$^{3}$)		& 0.3560$\cdot10^{6}$ 	$^a$ 	\\
			$C_{6}$ (cm$^{-1}$\AA$^{6}$)		& 0.4230226740$\cdot10^{8}$	  $^b$ 	\\
			$C_{8}$ (cm$^{-1}$\AA$^{8}$)	    & -0.2969367813$\cdot10^{10}$		$^b$\\
			diss.asympt. (cm$^{-1}$)				& 35481.745		 	$^c$			\\					
		\hline \noalign{\smallskip}			
			\multicolumn{2}{l}{adiabatic Born-Oppenheimer correction}	          \\ 			
			$v_{0}$  (cm$^{-1}$)  		&  1.4387\\	
		\hline \noalign{\smallskip}
			std. dev.  (cm$^{-1}$)             & 0.025   \\
			reduced std. dev. $\sigma$								&  0.97	\\
		\hline \noalign{\smallskip}
			\multicolumn{2}{l}{Derived constants}	          							\\ 
			R$_{e}$ (\AA)			    	& 3.0835						\\
			T$_{e}$ ($^{24}$Mg$_2$) (cm$^{-1}$)				& 26068.894(500)		\\	
			D$_{e}$ ($^{24}$Mg$_2$) (cm$^{-1}$)				& 9412.851(500)		 	\\	
		\hline \noalign{\smallskip}
		\end{tabular}
	  \begin{flushleft}
			 $^a$ calculated from lifetime \cite{Lundin}, $^b$ adjusted for continuous and differentiable 
			 transition at R$_i$ or R$_o$, $^c$ fixed value, see text \\
		\end{flushleft}
\end{table}

The same data set was used for a fit to Dunham parameters, which allow for a convenient calculation of an 
energy level E(v,J) from given parameters, because the level energies are expanded as a truncated power series
\begin{equation}
\label{Dunham}
  E(v,J)=T_0 +\sum_{i,k = 0} Y_{ik}(v+1/2)^i [J(J+1)-\Omega^2]^k
\end{equation}
where $v$ is the vibrational, J the rotational quantum number and $\Omega$ the projection of the total electronic
angular momentum on the molecular axis. Effects of Born-Oppenheimer corrections are included in this case in the parameter $\Delta_{00}$, such that 
\begin{equation}
[T_{0}+Y_{00}](^i\textrm{Mg}^j\textrm{Mg})=T_{00}(1+\Delta_{00}\frac{m_e}{\mu_{ij}}),
\end{equation}
where m$_e$ is the electron mass and $\mu$ the reduced mass of the isotope combination "i" and "j". This formalism is described in more detail e.g. in \cite{BOC-Dunham}. The Dunham representation fits the energy levels with similar quality as the potential fit does. The standard deviation is 0.025 cm$^{-1}$ for 14 parameters as listed in table \ref{tab2} and the reduced standard deviation becomes 1.0, not significantly larger than for the PEC fit.
\begin{table}
	\caption{Dunham parameters for the state \Astate~of $^{24}$Mg$_2$ for vibrational levels 1~$\le$~$v'$~$\le$~46. The uncertainties are purely derived from the statistics of the deviations in a linear fit.}
	\label{tab2}
		\begin{tabular}{ll}
		\hline \noalign{\smallskip}
		\hline \noalign{\smallskip}
			Parameter & Value\\
		\hline \noalign{\smallskip}
			T$_{00}$  (cm$^{-1}$)   &  26070.046(31)  \\
			Y$_{10}$  (cm$^{-1}$)	&  190.7571(14)		\\
			Y$_{20}$  (cm$^{-1}$)	&  -1.16567(18)	\\
			Y$_{30}$  (cm$^{-1}$)	&  0.28183(94)$\cdot10^{-2}$	\\
			Y$_{40}$  (cm$^{-1}$)	&  -0.2011(22)$\cdot10^{-4}$	\\
			Y$_{50}$  (cm$^{-1}$)	&  0.586(19)$\cdot10^{-7}$	\\
			Y$_{01}$  (cm$^{-1}$)	&  0.1480908(28) \\
			Y$_{11}$  (cm$^{-1}$)	&  -0.133164(33)$\cdot10^{-2}$\\
			Y$_{21}$  (cm$^{-1}$)	&  0.2887(16)$\cdot10^{-5}$		\\
			Y$_{31}$  (cm$^{-1}$)	&  -0.3234(27)$\cdot10^{-7}$	\\
			Y$_{02}$  (cm$^{-1}$)	&  -0.36085(78)$\cdot10^{-6}$	\\
			Y$_{12}$  (cm$^{-1}$)	&  0.808(70)$\cdot10^{-9}$\\
			Y$_{22}$  (cm$^{-1}$)	&  -0.352(15)$\cdot10^{-10}$		\\															
			$\Delta_{00}$  			& -1.023(50)	\\
			std. dev. (cm$^{-1}$)             & 0.025   \\
			red. std. dev. $\sigma$								&  1.0	\\										
		\hline \noalign{\smallskip}
			$[T_{0}$+Y$_{00}]$$(^{24}$Mg$_2$)  (cm$^{-1}$)	& 26068.835(40)		\\
		\hline \noalign{\smallskip}
		\end{tabular}
\end{table}
For both single state approaches the energy contributions of the Born-Oppenheimer correction to the \Astate~state are similar, the appropriate T parameter increases by about 0.1 cm$^{-1}$ when going from $^{24}$Mg$_2$ to $^{26}$Mg$_2$.

\subsection{Evaluation of the coupled system \Astate~+ \Bstate}
\label{coupled}

Figure \ref{fig:2} shows that observed level shifts start at about 2500 cm$^{-1}$ above the bottom of the \Astate~state. Comparing this with the electronic states in figure \ref{fig:1} it becomes very plausible that the coupling with the \Bstate~state is the reason for the level shifts. An interaction between a $^1\Pi_u$ and $^1\Sigma^+_u$ state will be caused by the non-diagonal L-uncoupling part of the rotational operator H$_{rot}$ in the Hamiltonian. It couples electronic states differing by one unit in $\Lambda$, the projection of the total orbital angular momentum on the internuclear axis. For the present case it can be expressed as \cite{Field}
\begin{eqnarray}
\label{nondiag}
  H_{rot, nondiag} =   \frac{\hbar^2}{2\mu R^2}( J^-L^+ +  J^+L^-).
\end{eqnarray}
If the expectation value of L$^+$ over the electronic space is defined as a function $\Xi$(R) of internuclear separation R, the matrix element of the operator above can be written in the Hund's case a basis of the two components $| ^1\Sigma^+_u,  J >$ and $| ^1\Pi_u,  J >$ as
\begin{eqnarray}
\label{nondiag1}
  < ^1\Pi_u,  J | H_{rot} | ^1\Sigma^+_u, J > =\Xi(R) \frac{\hbar^2}{2\mu R^2} \sqrt{J(J+1)}  
\end{eqnarray}
and $\Xi(R)$ will be represented as a power expansion in R like a potential, 
\begin{equation}
 \Xi(R) = <^1\Pi_u |L^+| ^1\Sigma^+_u >=D(R,R_c)\cdot\sum_{i=0} \xi_i (R-R_\xi)^i ,
\label{Rform}
\end{equation}
where R$_\xi$ is the expansion point in the interval of overlap of the two electronic wave functions thus normally close to R$_e$ of one of the potentials. In order to avoid unphysical behaviour of this function in regions of large R, where it is not well defined by data, the damping function D(R,R$_c$) of the form 
\begin{equation}
\label{damp}
  D(R, R_c)= \frac{1}{1+e^{\alpha(R-R_c)} }
\end{equation}
with $\alpha$ $>$ 0 is introduced to guarantee that $\Xi(R)$ goes to zero for R to infinity, because the electronic states \Astate~and \Bstate~correlate asymptotically to $^1$P+$^1$S and $^3$P+$^3$P, respectively, resulting to an expectation value zero for L$^+$. If $\Xi(R)$ can be assumed to be constant in the region of overlap of the vibrational wave functions, equation (\ref{nondiag1}) can be simplified in a rovibrational basis for the pair of vibrational levels $v$, $v'$ to
\begin{eqnarray}
\label{cform}
  < ^1\Pi_u, v,J | H_{rot} | ^1\Sigma^+_u, v',J > =  \nonumber \\
    \xi_0 < v,  J |\frac{\hbar^2}{2\mu R^2} | v', J > \sqrt{J(J+1)} ,
\end{eqnarray}
which will be used later for the simplified approach of a local deperturbation between the vibrational state $v$ of \Bstate~and $v'$ of \Astate.

The nondiagonal L-uncoupling is the only interaction taken into account here. It couples levels of the \Bstate~and \Astate~states with same e/f symmetry \cite{Field} and common total angular momentum J. Thus only the e levels of the \Bstate~state couple to the \Astate~levels, which leads to $\Lambda$-doubling in the $\Pi$ state. Unfortunately, we do not observe this due to missing spectral lines of this state. The coupled system is set up in the basis system defined above in a Fourier grid of R (see e.g. \cite{Lisdat}) using potential representations for states \Astate~and \Bstate~like those given by equation (\ref{uanal}) with appropriate extension for short and long range and the term from equation (\ref{nondiag1}) being non-diagonal in $\Lambda$ and the kinetic energy being non-diagonal in R. The eigenvalues are found by diagonalizing the large matrix in R with 1.58 \AA~$<$ R $<$ 11.3 \AA~and about 270 grid points giving a matrix dimension 540 x 540 of our two state model.

First attempts to use the \textit{ab initio} potential of the \Bstate~state modified to an "empirical" (e.g. \cite{empi}) potential by adding to it the difference of the experimental A state potential and the \textit{ab initio} A state potential directly were unsuccessful. This was mainly due to the fact that the constructed empirical potential is rather far off from the final one, and the computational demand by the Fourier grid code is large making response times fairly long.

Therefore, as a quicker and more flexible approach, a local deperturbation was preferred for constructing a sufficiently well approximated starting potential for the final Fourier grid fits. This approach uses the level energies of either state expanded into a Dunham series and puts these into the matrix instead of the potentials and of the kinetic energy. Then the Hamiltonian is represented in a basis $| ^1\Pi_u, v, J >$ and $| ^1\Sigma^+_u, v',  J >$. The required non-diagonal matrix elements according equation (\ref{cform}) with B$_{v,J,v',J}$ = $ < v,  J |\frac{\hbar^2}{2\mu R^2} | v', J >$ were calculated by applying the vibrational wave functions from a single state approximation for either state from RKR potentials using preliminary Dunham parameters. For values of total angular momentum J = 40, 60 and 80 tables were created, from which the desired values of B$_{v,J,v',J}$ were derived by interpolation. This is precise enough as the dependence of the matrix elements on the rotational quantum number J is small. 

A computer program was set up which for any observed level of the \Astate~state finds the closest interacting level of the \Bstate~state and determines the energies by diagonalizing the two by two matrix, which has the unperturbed energies on the diagonal and the non-diagonal element defined by equation (\ref{cform}). This restriction to closest neighbor coupling was used here only as a preliminary evaluation, but it is also interesting how far this significant simplification might be justified. For the \Astate~state a set of parameters like in table \ref{tab2} was used, but changes were allowed during the fit. However, for this approach only the data of the main isotopologue $^{24}$Mg$_2$ were used, thus no Born-Oppenheimer corrections appear. Finally 5409 (4955 unperturbed + 454 perturbed) levels of the \Astate~state were included in the fit, covering the range of vibrational levels 1 $\le$ $v'$ $\le$ 46 and of rotational quantum numbers 5 $\le$ J' $\le$ 87.

For the \Bstate~state, a set of preliminary T, Y$_{10}$, Y$_{20}$, Y$_{01}$, Y$_{11}$ and Y$_{02}$ was 
determined from the \textit{ab initio} potential, and the bundle of lines was overlaid on the plot in figure 
\ref{fig:2} for estimating the perturbing rovibrational structure of state \Bstate~by using E($v$,J)=T$_v$ + B$_v$[J(J+1)-1]. This graphical method allows quick refinement such that the crossings of the rovibrational ladders occur at the proper J, i.e. where the signs of the observed line shifts change. 

Iteratively, including step by step more vibrational levels of the \Astate~state, the parameters of both 
states were refined to finally include the whole data set. The vibrational numbering of the \Bstate~state was optimized for smallest $\chi^2$ of the total fit, while always checking, that the crossings of the energy 
ladders occur at the proper rotational quantum numbers.  The Dunham parameters of both states together with the coupling constant $\xi_0$ are collected in table \ref{tab3}.

\begin{table}
	\caption{Dunham parameters from the local deperturbation 
	 of the \Astate~and \Bstate~states of $^{24}$Mg$_2$. The parameters of the \Bstate~state were used for the calculation of the lines showing the \Bstate~levels in figure \ref{fig:2}. The uncertainties are determined from the locally linearised problem at the minimum of $\chi^2$.}
	\label{tab3}
		\begin{tabular}{lll}
		\hline \noalign{\smallskip}
		\hline \noalign{\smallskip}
			Par. & \Astate  & \Bstate\\
		\hline 
			$T_0$ (cm$^{-1}$) 	& 26068.8255(38)   &  27952.4(3)	\\
			$Y_{10}$ (cm$^{-1}$) 	& 190.7614(13)  & 249.54(6) 	\\
			$Y_{20}$ (cm$^{-1}$)	& -1.16626(17)  & -0.984(3) \\
			$Y_{30}$ (cm$^{-1}$)	& 0.28530(93)$\cdot10^{-2}$  &  \\			
			$Y_{40}$ (cm$^{-1}$)	& -0.2104(22)1$\cdot10^{-4}$  &	\\
			$Y_{50}$ (cm$^{-1}$)	& 0.678(19)$\cdot10^{-7}$  &	\\
			$Y_{01}$ (cm$^{-1}$) 	& 0.148094(26) &  0.17208(8)  \\ 
			$Y_{11}$ (cm$^{-1}$) 	& -0.133273(30)$\cdot10^{-2}$  & -0.1010(5)$\cdot10^{-2}$\\
			$Y_{21}$ (cm$^{-1}$)	& 0.2966(15)$\cdot10^{-5}$ &	\\
			$Y_{31}$ (cm$^{-1}$)	& -0.3369(26)$\cdot10^{-7}$ &	\\
			$Y_{02}$ (cm$^{-1}$)  & -0.36010(76)$\cdot10^{-6}$ &	-0.24(1)$\cdot10^{-6}$\\
			$Y_{12}$ (cm$^{-1}$) 	& 0.736(67)$\cdot10^{-9}$  &	\\
			$Y_{22}$ (cm$^{-1}$) 	& -0.340(14)$\cdot10^{-10}$ &	\\											
			$\xi_0$					&  0.636(4) \\
		\multicolumn{2}{l}{}   \\
			std. dev.(cm$^{-1}$)       & 0.026   \\
			red. std. dev. $\sigma$		  	&  1.0 \\
 		\hline \noalign{\smallskip}
		\end{tabular}
\end{table}
From this set of Dunham parameters RKR potentials were calculated with which the central parts of analytical starting potentials were constructed for the Fourier grid fit. The extension of the A state potential was done as in subsection \ref{constr}. The short range branch of the \Bstate~state was extended using the functional form of equation (\ref{rep}). The uppermost energy level involved of \Bstate~is still more than 2500 cm$^{-1}$ below the dissociation limit $^1$P + $^1$S, and even further away from the region of avoided crossing of the \Bstate~state with the (2)$^1\Pi_u$ state (see figure \ref{fig:1}). Therefore, the details of the potential extension for large R do not influence the energy levels involved in the fit of the central part of the \Bstate~state potential. Thus we 
ignored the avoided crossing to simplify the model and the \Bstate~state is extended by forcing a continuous and differentiable connection at R$_o$ using C$_6$ and C$_8$, and the dissociation asymptote at 44212.18 cm$^{-1}$ for the energy of the $^3$P + $^3$P pair referred to the bottom of the ground state PEC.

For the Fourier grid fit the data are treated in the same manner as for the approach in section \ref{single}, but now covering the full range of quantum numbers $v'$~$\le$~46 and J' of the \Astate~state with all 5844 level energies for the fit. The fit converged to the final potentials whose parameters are listed in table \ref{tab4}. The fit quality characterized by the reduced standard deviation $\sigma$ is as good as for the local deperturbation, but the lines of all isotopologues are included.

\begin{table}
	\caption{Potential parameters for unperturbed potential energy curves of the \Astate~and \Bstate~states of $^{24}$Mg$_2$ from the global deperturbation with the Fourier grid method.}
	\label{tab4}
		\begin{tabular}{lll}
		\hline \noalign{\smallskip}
		\hline \noalign{\smallskip}
			Parameter & \Astate   &  \Bstate  \\
    \hline
		  R$_i$ (\AA)	& 2.40   &  2.32 \\
		  R$_o$ (\AA)	&	7.00   &  3.98 \\
		\hline \noalign{\smallskip}
		\multicolumn{3}{l}{$R<R_{i}$} \\ 
			A (cm$^{-1}$) $^b$	      	& 0.232831962$\cdot10^5$ & 0.253221018$\cdot10^5$  	\\
			B (cm$^{-1}$\AA$^s$) $^b$		& 0.16881614$\cdot10^7$  & 0.12163567$\cdot10^7$\\
			s 					&  6  &  6 \\
		\hline \noalign{\smallskip}
		\multicolumn{3}{l}{ $R_{i}\leq R \leq R_{o}$ } 	  \\ 
			$R_{m}$ (\AA)		    	& 3.0810					  					&	 2.850	\\
			T$_m$ (cm$^{-1}$)     & 26068.928					  				&	 27950.50\\
			b						          & -0.50												  &	 -0.60 \\
			$a_{1}$   (cm$^{-1}$)	& -0.30868902012$\cdot10^{2}$ &	-0.1551075865$\cdot10^{3}$	\\
			$a_{2}$   (cm$^{-1}$)	& 0.15391080212$\cdot10^{5}$ &	0.1454901454$\cdot10^{5}$	\\
			$a_{3}$   (cm$^{-1}$)	& 0.91396765946$\cdot10^{4}$ &	0.1522044730$\cdot10^{5}$	\\
			$a_{4}$   (cm$^{-1}$)	& -0.12112855696$\cdot10^{4}$ &	0.8628269253$\cdot10^{4}$	\\
			$a_{5}$   (cm$^{-1}$)	& -0.16560723199$\cdot10^{5}$ &	-0.1658657656$\cdot10^{4}$	\\
			$a_{6}$   (cm$^{-1}$)	& 0.50027986049$\cdot10^{4}$ &	-0.3953560614$\cdot10^{4}$	\\
			$a_{7}$   (cm$^{-1}$)	& 0.87361885593$\cdot10^{5}$ &	\\
			$a_{8}$   (cm$^{-1}$)	& -0.69237989563$\cdot10^{5}$ &	\\
			$a_{9}$   (cm$^{-1}$)	& -0.40569186950$\cdot10^{6}$ &	\\
			$a_{10}$  (cm$^{-1}$)	& 0.21924570539$\cdot10^{6}$ &	\\
			$a_{11}$  (cm$^{-1}$)	& 0.99034939502$\cdot10^{6}$ &	\\
			$a_{12}$  (cm$^{-1}$)	& -0.40996854006$\cdot10^{6}$ &	\\
			$a_{13}$  (cm$^{-1}$)	& -0.13251321051$\cdot10^{7}$ &	\\
			$a_{14}$  (cm$^{-1}$)	& 0.31127669584$\cdot10^{6}$ &	\\
			$a_{15}$  (cm$^{-1}$)	& 0.75779011259$\cdot10^{6}$ &	\\
		\hline \noalign{\smallskip}
			\multicolumn{3}{l}{ $R_{o}<R$	}   \\ 
			C$_{3}$ (cm$^{-1}$\AA$^{3}$)$^a$		& 0.35600$\cdot10^{6}$ 	 & 	\\
			C$_{6}$ (cm$^{-1}$\AA$^{8}$)$^b$		& -0.4606515$\cdot10^{8}$ &	 0.1157708$\cdot10^{9}$\\
			C$_{8}$(cm$^{-1}$\AA$^{10}$)$^b$	  &  0.3134423$\cdot10^{10}$	 &	-0.1176734$\cdot10^{10}$ \\
			diss. (cm$^{-1}$)$^c$				& 35481.745	 & 	44212.180	\\		
		\hline \noalign{\smallskip}		
			\multicolumn{2}{l}{adiabatic Born-Oppenheimer correction}	          \\ 			
			$v_{0}$  (cm$^{-1}$)  		&  1.4465\\							
 		\hline \noalign{\smallskip}
			$\xi_0$   &  0.6389  & \\
			$\alpha$ (\AA$^{-1}$) & 2.5 & \\
			R$_c$(\AA)    & 5.0  & \\
		\hline \noalign{\smallskip}				
			std.dev.(cm$^{-1}$)       &  0.025  & \\
			$\sigma$			   &  0.97	 &\\
		\hline \noalign{\smallskip}
		\multicolumn{3}{l}{ derived constants }		\\ 
			$R_{e}$ (\AA)			      	& 3.0825(1) 	  &	2.856(2) 			\\
			$T_{e}$ (cm$^{-1}$)				& 26068.913(700) 		&	27950(10)    \\	
		\hline \noalign{\smallskip}
	\end{tabular}
	\flushleft
     	$^a$ from lifetime \cite{Lundin} \\
     	$^b$ adjusted for continuously differentiable transition at R$_o$ and R$_i$ \\
     	$^c$ from \cite{isv} and D$_e$ (X) = 430.472~cm$^{-1}$\cite{Xpaper}\\
	\end{table}	
\begin{figure}
\resizebox{0.5\textwidth}{!}{%
  \includegraphics{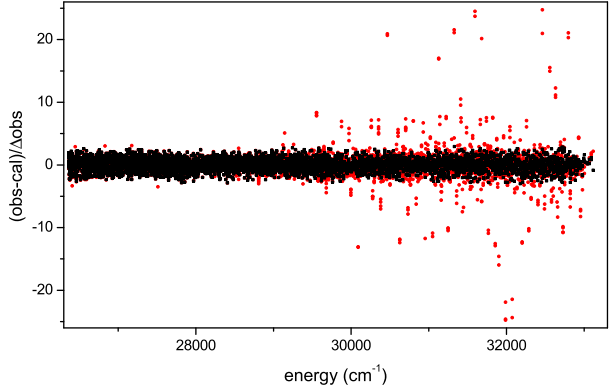} }
  \caption{(color online) Fit residuals normalized to the experimental uncertainty. Black dots represent 
   the result of the coupled channels fit. Red dots show the relative residuals using the result of the 
   single state approach for a simulation of all observations including the perturbed levels.} 
  \label{fig:3}
\end{figure}

We show in figure \ref{fig:3} the fit residuals normalized to the experimental uncertainty $\Delta$obs. Black dots give the result of the coupled channels fit. The distribution of the relative residuals was checked to be symmetric around zero. The average deviation agrees well with the anticipated experimental uncertainty. The red dots mark the normalized residuals created by a simulation with the result in table \ref{tab1} of the single state approach including the perturbed levels. Differences (obs-cal) up to 1 cm$^{-1}$ occur, showing in figure \ref{fig:3} magnitudes up to about 25. The collapse of the scatter of the red dots to the distribution of the black dots clearly demonstrates the success of describing the observations by a two-state model.

\section{Discussion and conclusion}
\label{disc}

All fit results tabulated above yield sets of parameters for the \Astate~ state of Mg$_2$, which allow in their range of applicability prediction of level energies to a mean precision of 0.025~cm$^{-1}$. Despite extending the range of vibrational levels compared to \cite{Balfour}, the uppermost levels are still more than 2500~cm$^{-1}$ below the asymptote Mg($^1S_0$)+Mg($^1P_1$), to which the \Astate~state correlates. Thus with the present set of data it is not possible to characterize more precisely the asymptotic behaviour of the PEC or to derive a C$_3$ coefficient from molecular spectra.

Through the identification of perturbed levels of the \Astate~ we evaluated the perturbing state to reasonable accuracy by Dunham parameters and by a PEC in a coupled channels fit employing the Fourier grid method. The perturber is assumed to be the \Bstate~state according to the \textit{ab initio} calculations \cite{ai-pots,Old-ai}. Dunham parameters and the PEC, however, should be taken with some care. As there are no direct observations of transitions 
to the \Bstate, it is only possible to find a vibrational numbering which is consistent with the present experimental data situation. So in future, if better data and extensions to transitions \Bstate~- \Xstate~are available, a revision might be necessary, perhaps yielding a modified vibrational numbering. We have used the present PEC for the \Bstate~and calculated Franck-Condon factors (FCF) in order to check, if our failure of observing direct transitions between \Xstate~and \Bstate~is supported by the FCFs. Indeed, the largest FCFs, yet less than 0.08, are calculated for a (26-0) band, which would appear around 33750~cm$^{-1}$, beyond the region which was included in the present evaluation. Weaker bands, e.g. FCF $\approx$ 0.02, which could be identified for the A-X system, would appear for \Bstate~- \Xstate~at about 32000~cm$^{-1}$, however, were not observed. This might be due to the substantially smaller transition dipole moment for the \Bstate~- \Xstate~transition compared to the \Astate~- \Xstate~transition for short internuclear distances \cite{ai-pots}. Such behavior of the transition dipole seems quite plausible if one takes the 3s3p + 3s3p configuration as the dominant contribution for \Bstate~as suggested from the electronic structure in figure \ref{fig:1} for large R.

\begin{figure}
\resizebox{0.5\textwidth}{!}{%
  \includegraphics{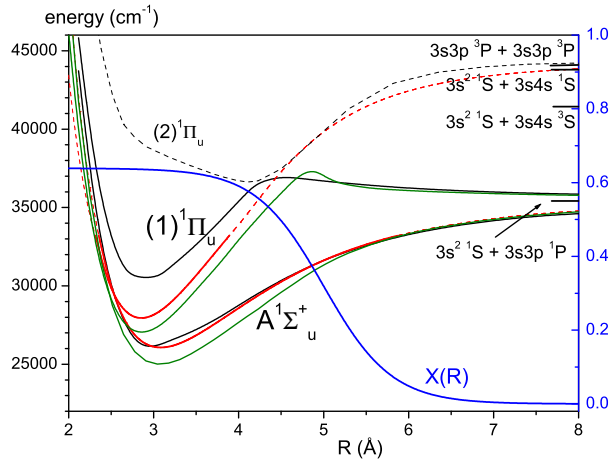} }
  \caption{(color online) Potential energy curves from experiment compared to \textit{ab initio}
  PECs. Red lines: Dotted: full numeric model potential with extensions as used in Fourier Grid fits, 
  thick full lines: range of potentials supported by experimental data. Blue: coupling function.} 
  \label{fig:4}
\end{figure}

The experimentally determined parts of PECs of both excited states are shown in red in figure \ref{fig:4}, together with the resulting coupling function $\Xi$(R) (blue). The coupling function turned out to be by no means uniquely determined by the fits. We tried various shapes, which all gave the same fit quality. They all intersect at the radius R$_{cr}$ = 2.52~\AA~ of the crossing of the PECs, that means they have the same value there. Obviously the coupling function is well determined by the data only around this point R$_{cr}$. We decided to give in table \ref{tab4} the simplest form, with the constant $\xi_0$ and the damping function. The value $\xi_0$ of the coupling function corresponds well to the result in table \ref{tab3} of the local fit. Its magnitude is of order one. This is quite reasonable, a change from $\sigma$ to $\pi$ orbital for a p electron would give $\sqrt2$, \cite{Field}, however, the configurations of the interacting states here differ probably by more than one electronic orbital reducing the coupling.

Comparing the performance of the models used for the coupled state description, local deperturbation versus global one, both work in this particular case similarly well. This happens probably due to the special situation, that the coupling is localized close to the steep repulsive branch.

	\begin{table}
	\caption{Characteristic parameters of the \Astate~and \Bstate~excited states of $^{24}$Mg$_2$
	 from the present experimental study and \textit{ab initio} calculation \cite{ai-pots}. T$_e$ 
	 is referred to minimum of the ground state.}
	\label{tab5}
		\begin{tabular}{lllll}
	  &  T$_e$(cm$^{-1}$)  &  D$_e$(cm$^{-1}$)  &   R$_e$(\AA)  & $\omega_e$(cm$^{-1}$) \\
		\hline \noalign{\smallskip}
		\hline \noalign{\smallskip}
		\Astate & & & & \\
		 this work &   26068.9   &   9414   &  3.0825  &  190.76  \\
		 \cite{ai-pots} & 26097  &  9427  &  3.04  &  19.8  \\		
		 \cite{Old-ai} & 25005  &  10476  &  3.10  &  191.5  \\
		\Bstate & & & & \\
		 this work &   27950  &  7532$^a$  &  2.856   &  249.5 \\ 
		 \cite{ai-pots} & 30518 &    4964  &  2.91  &  246.3  \\		
		 \cite{Old-ai} & 26987 &    8494  &  2.83  &  252.0  \\
		\hline \noalign{\smallskip}
	\end{tabular}
	\flushleft
	$^a$ referred to asymptote Mg($^1$P$_1$) + Mg($^1$S$_0$) at 35481.745~cm$^{-1}$.
	\end{table}
	
In figure \ref{fig:4} we also show the PECs from \textit{ab initio} calculations, the full black curves from \cite{ai-pots} and the full green curves from \cite{Old-ai}. While the PEC for the \Astate~state calculated by \cite{ai-pots} is rather close to the experimental PEC, the \Bstate~curve is about 2500~cm$^{-1}$ higher in energy than the experimental one. The PECs by \cite{Old-ai} are both lower in energy than their experimental counterpart, by about 1060~cm$^{-1}$ for the \Astate~and about 960~cm$^{-1}$ for the \Bstate. For a comparison between experiment and theory we list few conventional molecular parameters in table \ref{tab5} (the vibrational frequencies $\omega_e$ for the PECs by \cite{ai-pots} were determined from the eigenvalues of their rotationless potentials). In contrast to the partly large deviations of T$_e$, which result also in large differences in the bond strength characterized by 
D$_e$, the equilibrium internuclear distances R$_e$ and the vibrational frequencies $\omega_e$ are close to the experimental values. Because of the significant differences in the electronic energies of the two \textit{ab initio} approaches it is quite important to analyze the origin of such differences.

\begin{figure}
\resizebox{0.5\textwidth}{!}{%
  \includegraphics{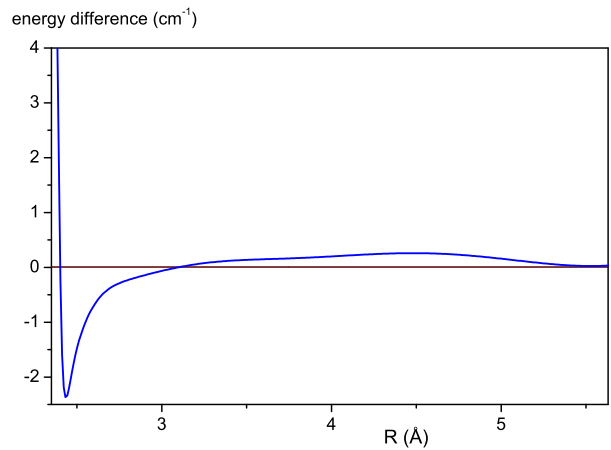} }
  \caption{(color online) Difference of PEC by coupled channel fit - PEC from single state fit. Only the R-interval supported by the present data is shown.} 
  \label{fig:5}
\end{figure}

Finally, in figure \ref{fig:5} we compare the PECs of the state \Astate~from the single state fit and the coupled channels fit. The difference is small for the major part of the interval. This means that apparently the choice of levels, which were regarded as being unperturbed, already determines the potential quite well for further applications. The perturbations affect mostly the short range branch, where the perturbing state is located. The differences of the PECs, however, translate to much smaller differences of eigenvalues.

Above the region of levels v$_A$ $\le$ 46 discussed here the assignment could not be extended with the perturbation structure given in figure \ref{fig:2}. Perturbations appear more often. This might hint to additional influence of other electronic states than the \Bstate~state on the \Astate~state, like e.g. $\Sigma$ states developping from other asymptotes as indicated on the right of figure \ref{fig:4}, giving rise to further perturbations.

In conclusion, for both electronic states we give potential energy curves and the coupling function, which describe 
the presently available data within a standard deviation of 0.025~cm$^{-1}$. The experimental PECs differ significantly in energy from the \textit{ab initio} potentials, while equilibrium internuclear distances and vibrational frequencies fit well.

We hope that these experimental findings will trigger new theoretical effort and help to reduce the discrepancy between experimental and theoretical PECs for the alkaline earth metals.

\section{Acknowledgement}
We thank the Warsaw theory group headed by Robert Moszynski for providing their latest \textit{ab initio} potential energy curves. The work has been carried out with support from the Centre of Excellence QUEST.  S.R. gratefully acknowledges financial support by E. Rasel and E.T. support from the Minister of Science and Culture of Lower Saxony, Germany, by providing a Niedersachsenprofessur.

\end{document}